\documentclass[aps,pr,preprint,groupedaddress,nofootinbib,byrevtex]{revtex4-1}
\usepackage{amsmath,amssymb}
\usepackage{graphicx}
\usepackage{dcolumn}
\usepackage{bm}
\usepackage{color}
\usepackage{ulem}
\usepackage{float}
\usepackage{braket}

\usepackage{enumerate}

\begin{document}
\title{  New  class of quantum transitions exhibiting large-scale  intercorrelations : Color of the sky } 

\author{Kenzo Ishikawa${}^{1,2}$ and {Masaki Takesada${}^{1}$}}
\affiliation{{${}^1$Department of Physics, Faculty of Science, Hokkaido
University, Sapporo 060-0810, Japan}, \\
{${}^2$Natural Science center, Keio University, Yokohama 223-8521, Japan},\\ }  
\date{\today}

\begin{abstract}

The absolute value  of the transition probability of the Rayleigh scattering  is computed for the first time and applied  to the scattering of solar light with molecule in the atmosphere and to the laser scattering with nanoparticles.   The probability has  a  new contribution of unique properties from long-range correlations specific to the quantum mechanics.  These  
 reveal  intercorrelated  reactions of unusual properties.  The magnitude is sufficient to  resolve   longstanding puzzle on   diffusion lights in the sky and  anomalous photon spectrum in laser experimens.    
The earth's albedo  from the new calculations  on Rayleigh scattering agrees well with observations  with satelites.

\end{abstract}
\maketitle
{\bf Introduction}

Light quanta   emitted from solar surface of low charge density are  wave packets of large coherence lengths. Quantum mechanical  Rayleigh scattering  of these lights with molecules in the atmosphere, which  is a fundamental process to the earth enviorement, is found  to deviate substantially 
from the classical picture owing to a long-range correlation of intercorrelated states.\cite{Rayleigh} The new results fill the gap of  the previous formulae obtained from the short range correlations with observations, and  are in agreement with  the brighness of the sky that  are hard to understand from the standard scattering theory.    The present formalism, furthermore,  can account for  the anomalous energy spctrum of scattered light of  recent  experiment.

 Absolute values of  transition probabilities depend on the time interval between the initial and final states, and  are expressed by rates and  other constants.  We classify the probability into the following categories. The first class and second class.   The  probability per unit of time, rate,  is  the first class 
  \cite{Landau-Lifshitz,Reed-Simon,Kato},  and  the  constant in time is the second class. The first class  attributes to an  expectation value of an operator with a single state, whereas the second class   attributes to  a pair of states,  which  reveals the long-range intercorrelated properties.   The long-range correlation was proposed first by Einstein-Podolsky-Rosen(EPR), and confirmed in measurements later \cite{EPR,Aspet}. Owing to its long-range nature, the second class quantities   depend on experimental situations, and are not extracted easily from the analyses which focus the first class quantity.   
   
     In new formula,  the scattering probability is the sum of the first and second classes,
      \begin{eqnarray}
d  P&=& {\kappa^2} d \Omega e^{  -R  -{\sigma_s} ({\delta \vec P})^2}  \left(   \theta(T_{int})   e^{-\sigma_t(\delta \omega )^2}+  \frac{ \sigma_t}{2\pi}  e^{-\frac{(T_{int}-T_0)^2}{\sigma_t}} \frac{1}{(T_{int}-T_0)^2+({\sigma_t}  \delta \omega )^2} \right), \label{Probability_{Rayleigh_s}}
\end{eqnarray}
where   the first class  agrees with the standard one computed with plane waves and the second class is  new.    The text   presents its derivation,  meanings of each term, and how this explains  the blue sky and anomalous photon spectrum in the laser experiment.  
   

The present formalism  fills a deficiency in  the standard  plane-wave formalism, i.e.,  unique value of the probability is not provided from  the amplidudes  proportional to the Dirac delta function. \cite{Schwartz}
 Although  the probability for short distance fluctuations is computed correctly   by combining the modified interaction $e^{-\epsilon |t|}H_{int}$ with the renormalization for ultraviolet divergenses \cite{Weinberg,Abrikosov},   long distance fluctuations are incorrectly  dropped   as was pointed out  in the Feynman's first paper  of the Quantum Electrodynamics (QED).  \cite{Feynman,Schwinger} Wave packet formalism  provides  the full probability of whole phenomena   caused by  the interaction $H_{int}$. 
 
  Wave packet $|{\vec P}, {\vec  X} \rangle $ expresses   a normalized superposition of free waves  around a  momentum $\vec P$ and a position ${\vec X}$   not only satisfies   the free Schr\"{o}dinger  equation but also forms  a complete set \cite{ishikawa-shimomura-PTP}.
For a scalar   field,  a wave packet of  a size $\sigma_{M}$, an energy $E(\vec P_M )$, a momentum     $\vec P_M $, and a position $\vec X$  is proportional to 
$e^{-\frac{1}{2 \sigma_M}  ({\vec x}-{\vec X}_M-{\vec V}_M(t-T_M))^2-i(\tilde E({\vec P}_M) (t-T_M )-{\vec P}_M\cdot({\vec x}-{\vec X}_M )) }$, with ${\vec V}_M=\frac{{\vec P}_M}{E({\vec P}_M)}$,
%
in the  Gaussian wave packets. Other fields have similar expressions and are expressed in Supplements.  The wave packet size $\sigma_M$ is a  parameter that depends on envioremental circomstances  and  plays important roles in the second class quantities. Its size is estimated in \cite{ishikawa-jinnouchi}.

 Using the wave packets in the interaction picture, \cite{Ishikawa-Nishio, Ishikawa_1}   we solve  the Schr\"{o}dinger  equation  and obtain     
  transition  probability of an  initial state    $\psi_{in,\alpha}(t)$   at  $t=0$  to a  final state  $\psi_{out,\beta}$   at $t=T$ for a coupling strength $g$,   $P_{\beta \alpha}(g,T)=|\langle  \psi_{out,\beta},T|\psi_{in,\alpha},0 \rangle|^2 $.
$P_{\beta,\alpha}(g,T)$, satisfies 
$0 \leq P_{\beta,\alpha}(g,T) \leq 1$  and a manifest unitarity, $ \sum_{\beta} P_{\beta,\alpha}(g,T)=1$. 
 Contraly to divergent probability of  plane waves    at a finite $T$  \cite{Stuckelberg} \cite{Gell-Mann-Low}, that is expressed by 
$P_{\beta,\alpha}(T)=\Gamma_{\beta,\alpha} T +P^{(d)}_{\beta \alpha} $
for $P_{\beta \alpha}(T) \ll 1$.   $ \Gamma_{\beta \alpha}$ was derived in atomic tranitions and  extended  to wide transitions \cite{Dirac,Fermi}.    
Because $P^{(d)}_{\beta \alpha}$ is   not only independent of   $\Gamma_{\beta \alpha}$ but governed by new scale, this   is not negligible sometimes even for large  $T$.   
The physical quantities in $\Gamma_{\beta \alpha}$ are the first class and measured through the shifts of the energies.   $P^{(d)}_{\beta\alpha}$   is the second class. A finite $P^{(d)}_{\beta \alpha}$   is not in   contradiction with the  agreements of the theory  with the experiments for first class quantities including the  precision tests of QED.  \cite{Ishikawa-Tobita_1,Ishikawa-Tobita_2,Ishikawa-Tajima-Tobita,Ishikawa-Oda,Ishikawa-Nishiwaki-Oda, Ishikawa-Nishiwaki-Oda_2}.
    Possible signals  of $P^{(d)}$ in experiments  have been  barried 
  in the background due to unique properties. 
  This paper  presents  those phenomena that are lead from the second class probability,  $P^{(d)}$.


Hereafter we apply the wave-packet formalism to the Rayleigh scattering and other light scattering  and present   absolute values of probabilities of  these processes.  Because  the first class quantities  are understood well and in  agreement with   the standard formula in the literature and with experiments, we focus on the second class quantities. It is shown  that
 the new contributions   have sizable magnitude  and unique properties to  resolve puzzles on colors of the sky and the earth albedo.

{ \bf Reighlei scattering}

The Rayleigh scattering is a light scattering   with a molecule or  atom of  much smaller sizes  than the wave length.  
A transition amplitude  in the first order in a coupling constant for an initial state of a  light  of  momentum ${\vec p}_{\gamma_1}$ at a  spatial position ${\vec X}_{\gamma_1}$ at a time $T_0$ and a wave packet size $\sigma_{\gamma_1}$ and an atom at rest $\vec P_M=0$ at  a position $ {\vec X}_M$ at a time $T_0$ and a wave packet size $\sigma_M$, and  a final state of a light of  ${\vec p}_{\gamma_2}$ at ${\vec X}_{\gamma_2}$ at a time $T_1$ and wave packet size $\sigma_{\gamma_2}$ for a dipole interaction  between the matter field $\phi_M(x)$ and the electric field $F_{0i}(x)$,  $H_{int}=g \int d^3 x F_{0 i}(x)^2 \phi_M(x)^2$,  is expressed by a sum of two terms 
\begin{eqnarray}
S&=&i  \kappa \left(\prod_{A=1}^4(\pi \sigma_A)^{-3/4} \frac{1}{\sqrt{ 2E_A}} \right) e^{-\frac{\sigma_s}{2}( \delta \vec P)^2-\frac{R}{2} +i\theta_0}  (2 \pi \sigma_s)^{3/2}(2 \pi \sigma_t)^{1/2} (A+B), \label{amplitude_s2}   
\end{eqnarray}
using a spatial size  $\sigma_s$ and  a temporal  size $\sigma_t$ of combined wave packets. This amplitude is valid in  Rayleigh scattering with atom also. See reference  \cite{Rayleigh_{exp} } for Rayleigh  scattering of  an atom.   The detailed forms of parameters,   derivations of amplitudes, and computations of the probabilities   are presented  in Supplements.  $\kappa$ is proportional to  $g$ and  $A$ and $B$ are the bulk and boundary contributions,  which  depend upon the momenta and positions of the initial and final states. These  are given as 
\begin{eqnarray}
 A&=& \frac{1}{2} e^{\frac{\sigma_t}{2} (i \delta \omega )^2 } \left[sgn(T_{int}-T_0 )- sgn(T_{int}-T_1) \right], \nonumber \\
B&=&- \frac{1}{2}     e^{-\frac{(T_{int}-T_0)^2}{2 \sigma_t} +i(\delta  \omega )(T_{int}-T_0)} \sqrt{\frac{2 \sigma_t}{\pi}}  \frac{1}{T_{int}-T_0- i\sigma_t \delta  \omega } \nonumber \\
 &+& \frac{1}{2}   e^{-\frac{(T_{int}-T_1)^2}{2 \sigma_t} +i(\delta   \omega)   (T_{int}-T_1)} \sqrt{\frac{2 \sigma_t}{\pi}} \frac{1}{T_{int}-T_1- i\sigma_t \delta  \omega  }, \label{amplitudesAB}
\end{eqnarray}
where $T_{int}$ represents  the  time when the initial and final states intersect and $R$ is the trajectory factor.  These are functions of  $\vec {\tilde X}_{\gamma_1}=\vec X_{\gamma_1}-\vec V_{\gamma_1}T_1,  \vec {\tilde X}_{\gamma_2}=\vec X_{\gamma_2}-\vec V_{\gamma_2}T_2$ and explicit forms will be shown later.  The coupling strength $g$ is proportional to the fine structure constant $\alpha$. Because  its square is much smaller than $g$, higher order terms are much smaller than this order and are ignored in the present paper. \cite{higher_order}

Ignoring an interference term between the zeroth order  and  first order terms, and the one between $A$ and $B$  from reasons mentioned in the appendix, we have a probability distribution   
\begin{eqnarray}
d  P&=& {\kappa^2} \frac{1}{2E_1}  
  \frac{d^3P_{\gamma_2}}{(2\pi)^3 2E_{\gamma_2}} (2\pi)^4 \left( \sqrt{\frac{\sigma_t}{\pi}}\right)(\frac{\sigma_s}{\pi})^{3/2}  \frac{d^3X_{\gamma_2}}{( \pi \sigma_{\gamma})^{3/2}} e^{  -R  -{\sigma_s} ({\delta \vec P})^2} \nonumber \\
& &\left(   \theta(T_{int},T_0,T_1)   e^{-\sigma_t(\delta \omega )^2}+  \frac{ \sigma_t}{2\pi}  e^{-\frac{(T_{int}-T_0)^2}{\sigma_t}} \frac{1}{(T_{int}-T_0)^2+({\sigma_t}  \delta \omega )^2} \right),\label{Probability_{Rayleigh}} 
\end{eqnarray}
where $\theta(T_{int},T_0,T_1)=\theta(T_{int}-T_0) \theta(T_1-T_{int})  $ and
$\kappa^2=g^2 \frac{1}{2}p_{\gamma_1}^2 p_{\gamma_2}^2 (1+\cos \theta^2)$,  
in the natural unit $c=1,\hbar=1$. The first term in the right-hand side  provides  the transition rate, which  is exponentially suppressed with $\delta \omega $. The  second term is  $P^{(d)}$ of  power suppressed and sizable at $\frac{|T_1-T_0|}{\sqrt \sigma_t} \leq 1  $ and  negligible for $\frac{T_1-T_0}{ \sqrt{\sigma_t}} \gg 1 $.  Two terms  in Eq.$(\ref{Probability_{Rayleigh}})$  behave  differently    on the space-time positions as well as the momenta.  The first term   agrees with  the  standard  expression of Rayleigh scattering  peaked in a narrow energy window, whereas the second term decreases slowly in power  with respect to  the energy difference. Both terms depend on the positions.  Notably the second class quantity has sizable magnitude and  distinctive propereties.  

{\bf Magnitude of the second class quantities} 

 We note that the probability depends on the positions through  variable $ \tilde {\vec X}=\vec X-\vec V T$. 
Accordingly,  observed quantities at one position  in finite time interval  is equivalent to those at finite interval of positions at one time. 
 Exponent $R$ in Eq.$(\ref{Probability_{Rayleigh}})$   is reduced   for large $\sigma_{\gamma}$ to   
\begin{eqnarray}
\frac{ R}{2}
 =\frac{1}{2 \sigma_{\gamma}} [((\tilde X_{\gamma_1}-\vec X_M)_T^2+(\tilde X_{\gamma_2}-\vec X_M)_T^2+\frac{1}{2}((\tilde X_{\gamma_1} - X_M)_L -(\tilde X_{\gamma_2}- X_M)_L)^2]. \label{position}
\end{eqnarray} 
Because at $ |\tilde X_{\gamma_1}-\vec X_M| \ll \sqrt \sigma_{\gamma}$, $e^{-R} \sim  1$,  the probability is uniform in  $ \tilde X_{\gamma_1}-\vec X_M $. The scattering into the cone theorem of the standard formalism \cite{Scattering_ into_Cone} does not hold in this situation.
Integration over the transversal momentum is proportional to area of the surface,
\begin{eqnarray}
 \int   d^2 {X_{\gamma_2}}_T=S.
\end{eqnarray}
Integration  over the longitudinal position is converted to the intersection time, $T_{int}$.  Those of bulk term $|A|^2$ for  $\omega=V_{\gamma}(P_{\gamma_2}-P_{\gamma_1})$ in the spatial region ${X_{\gamma_2}}_L <   L_0$ is given by
\begin{eqnarray}
2 \pi \int d X_{\gamma_2 L} P_{\gamma_2} dP_{\gamma_2} |A|^2 =2 \pi\frac{P_{\gamma_1}}{V_{\gamma}} L_0  \frac{\sqrt \pi}{\sqrt \sigma_t} 
\end{eqnarray}
and is proportional to the interval.
The integration over boundary term in the extreme case $ \sqrt \sigma_t >> V_{\gamma}(T_1-T_0)$ or $ \sqrt \sigma_t<<V_{\gamma}(T_1-T_0)$ is given by, 
\begin{eqnarray}
2 \pi \int dX_{\gamma_2 L}  P_{\gamma_2} dP_{\gamma_2} |B|^2&=&2 \pi\frac{P_{\gamma_1}}{V_{\gamma}}\frac{1}{4} \log \frac{\omega_{max}}{\omega_{mini}} ( \sqrt \sigma_t >> V_{\gamma}(T_1-T_0)) \nonumber \\ 
&=& 2 \pi\frac{\sqrt {2 \pi}}{4} \frac{P_{\gamma_1}}{V_{\gamma}}  ( \sqrt \sigma_t << V_{\gamma}(T_1-T_0)).
 \end{eqnarray}
where the log factor in the right-hand side comes from the integration over the momentum and is expected around order $10$.
The ratio of the boundary term over the bulk term is
\begin{eqnarray}
\frac{\int dX_{\gamma_2 L}  P_{\gamma_2} dP_{\gamma_2} | B|^2}{\int dX_{\gamma_2 L}  P_{\gamma_2} dP_{\gamma_2} | A|^2}
 &=&
\frac{ \sqrt{ \sigma_t}}{4L_0  } \log{\frac{\omega_{max}}{\omega_{min}}} \gg 1 (~~\text{large $\sigma_t$ })  \nonumber \\
&=&
\frac{ \sqrt{2 \sigma_t}}{4 L_0  } \ll 1 (~~\text{small  $\sigma_t$ }). \label{ratio}
\end{eqnarray}
Because the ratio of second class quantities  over  first class quantities  is proportional to  $\frac{\sqrt \sigma_t}{L_0}$, the second class quantities is sizable  in the processes of large $\sigma_t$.


{\bf  Diffsed solar lights}

Solar lights  are wave-packets of large sizes, of an order $100$ kM, \cite{allen, withbroe}. The  scattered waves of the same wave-packet sizes are observable quantities and  sizes   
$\sigma_{\gamma_1}=\sigma_{\gamma_2} =\sigma_{\gamma} \gg \sigma_M,~~~
\sigma_s=\frac{\sigma_M}{2} \ll \sigma_{\gamma},
$ are studied  hereafter. 
Kinematical variables are expressed by 
${\vec V}_0=\frac{\sigma_M}{2\sigma_{\gamma}}({\vec V}_{\gamma_1}+{\vec V}_{\gamma_2 } ) =O(\frac{\sigma_M}{\sigma_{\gamma}})$, 
$\delta \omega= V_{\gamma}(P_{\gamma_1}- P_{\gamma_2})+O(\frac{\sigma_M}{\sigma_{\gamma}})$,
$ (\delta {\vec P})^2=(P_{\gamma_2}- P_{\gamma_1} \cos \theta)^2+P_{\gamma_1}^2(1- \cos^2 \theta)$, 
$\frac{1}{2\sigma_t}=\frac{1}{\sigma_{\gamma}} V_{\gamma}^2$
and $ T_{int}=-\frac{  ( {  \tilde X}_{\gamma_1} -\vec X_M)_L +(  {\tilde X}_{\gamma_2}-\vec X_M)_L }{ 2 V_{\gamma_1}  }$ 
where $\vec V_L$ and $\vec V_T$ are the longitudidal and transversal componentes of the vector $\vec V$.  The energy and momentum dependence  is governed by 
\begin{eqnarray}
& &\frac{\sigma_t}{2} (\delta \omega)^2+\frac{\sigma_s}{2} (\delta \vec P)^2 =+\frac{\sigma_{\gamma}}{4}  (P_{\gamma_1}- P_{\gamma_2})^2   
\end{eqnarray}
and the final energy agrees with the initial energy.

{\bf Evaluation of  the probability} 

We evaluate the strength of scattered light.   Eq.$(\ref{position})$ shows that  the strength  at the position $\vec X_{\gamma_2}$  does not depend upon  the molecule's position ${\vec X}_M$ as far as the condition $| {\vec X}_{\gamma_2}-\vec V_{\gamma_2} T_2||  \ll \sqrt{ \sigma_{\gamma}} $, is satisfied. This is a typical behavior of  quantum mechanical processes where the scattered light behaves as one quantum. Accordingly,  the light scatterered by molecules at a height $h$ has a uniform strength at a height $h$ of $ h \ll h_0$,where    $h_0 \approx \sqrt{\sigma_{\gamma}}$.   Incoherent sum of lights at a hight $h \leq h_0, h_0 \approx \sqrt {\sigma_{\gamma}}$ reveals the uniform strength   even though  the molecule density decreases rapidly with the height. This property  is specific to the quantum mechanical proceses caused by overlapp of  the large light wave packets.  

In a classical  physics,    a strength of scattered light emmited by a molecule    at an altitude   $h$  is  proportional to the  molecule density 
there $n(h)$, because  electromagnetic waves at each space-time position   are observable. When the density $n(h) $ decreases steeply with $h$,  the  strength of the waves decreases steeply with the height. The strength of their superpositions decreases steeply also.

\begin{figure}
\begin{center}
\includegraphics[width=120mm]{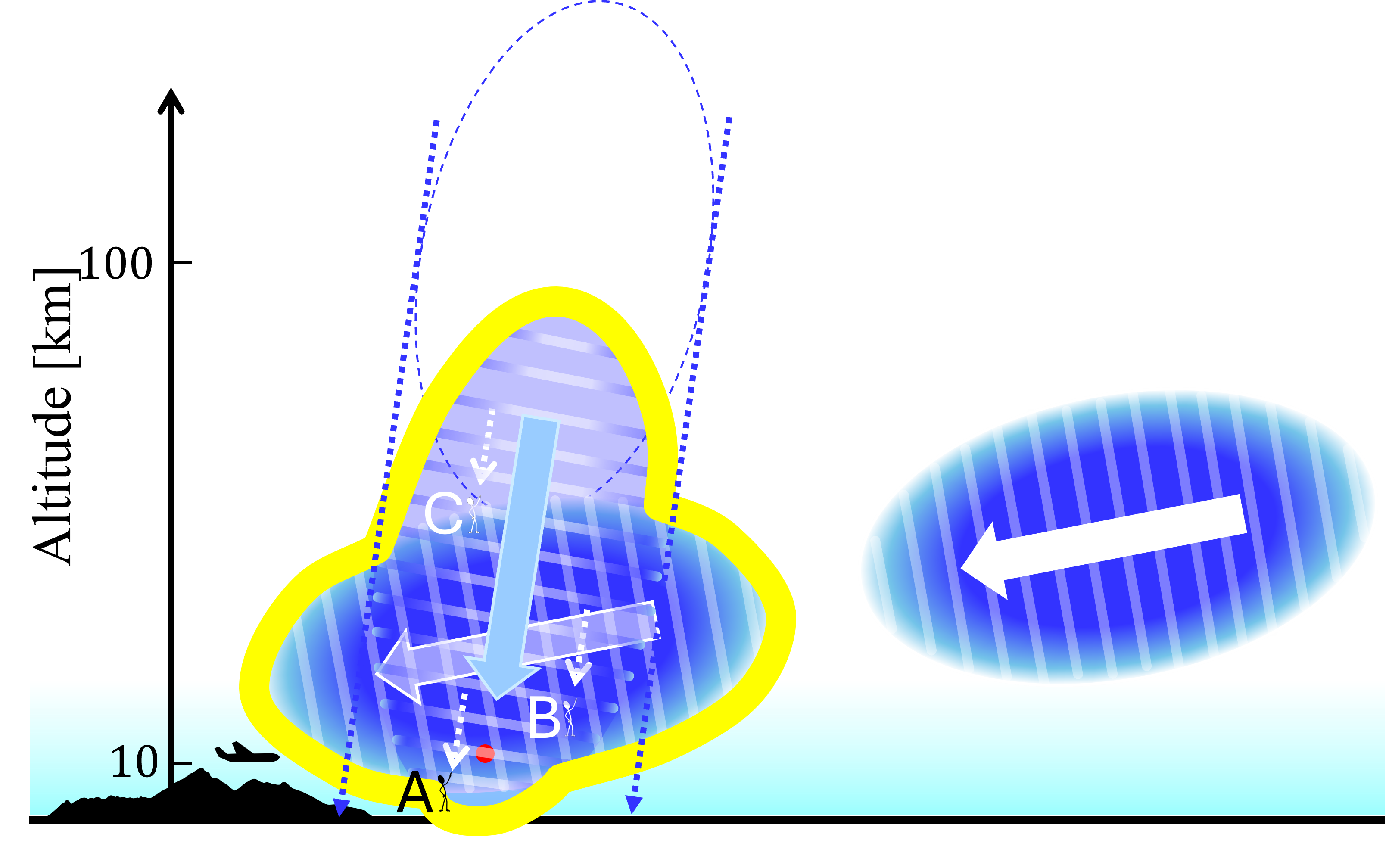} 
\caption{ Space-time view of Rayleigh scattering of a sun light of large wave-packet of an ellipse with a molecule. A wave packet of photon with white arrow is coming from the right, and scattered. Scattered wave  of wave vector of  blue arrow is observed by a person at A  as if that comes from the top.   Arrows show photon  momena and directions.  A small red dot shows a molecule of a scatterer. Interaction  occurs at a spatial area where the wave packets overlaps with the molecule. A scattered light can be observed at an altitude   10-20[km] 
even though  the molecule is near the earth surface, if the sizes are longer than  $R=30$ k.
}
\end{center}
\end{figure}

The size of matter wave packet,  $\sigma_M$,  is  estimated from the length that the particle keeps its coherence. The mean free path in the material or a coherence length of a solution  in the external fields show the sizes.  In the atmosphere of the earth, there is the gravity. A  kinetic energy and  the gravity potential energy agree at  $l=5 \times 10^{-7}$ M. A coherence length would be around this value, and $l= 10^{-7}$ M is used in the following evaluations. 
For the sun lights, $\sigma_{\gamma}$ is around $100^2$ kM$^2$, and  the one with  the solid material is around  $\sigma_{\gamma}= \pi r_m^2=  5 \times 10^{-12} m^2$.
In gas, density is low. Mean free path of lights in the atmosphere at the ground, $h=0$, is $20 \times 10^3$ M and longer at a higher altitude  and that from the sun is of the order $50\times 10^3$ M \cite{allen,withbroe}. Typical values of the wave packets are
\begin{eqnarray}
\sigma_s= 1.6 \times 10^{-14} \text{ M}^2, \sigma_{t}=3.9  \times 10^9 \text{ M}^2.
\end{eqnarray}
In  the scattering at a high altitude,   $\sigma_s$ is small but $\sigma_t$ is extremely large.  The second term in  Eq.$(\ref{Probability_{Rayleigh}}  )$  is sizable and does not depend on $\vec X_{\gamma}$  in a height $h< 50\times 10^3$ M.  In this  height,  the  strength of the scattered light is proportional to the number of the atoms in this region. These behaviors are characteristic features of the large wave packet  and distinct from the classical Rayleigh Scattering. This agrees  with the observation of  a diffused solar light in the sky. The classical Rayleigh scattering,  the strength   is  determined locally by scatterers and  total probability at one position  is proportional to their  density there. 

{ \bf Albedo of the Earth} 

We note that at the troposphere $h=0-10$ kM, there are cloud activity, and at  tratsphere $ h=10-50 kM$, there is no cloud activity.
So we evaluate the probability for the case   $L_0=10$ kM, and we have ratio of the second class over the first class from Eq.$( \ref{ratio})$
\begin{eqnarray}
\frac{ \sqrt{ \sigma_t}}{4 L_0  } \log{\frac{\omega_{max}}{\omega_{min}}}   \sim 2.5  \times O(1)  
\end{eqnarray}
with $\sqrt \sigma_t=100$kM and  $L_0=10$kM. The log factor would be between 1 and 10.
 The absolute value  of the probability   Eq.$(\ref{Probability_{Rayleigh}}  )$  is not only much larger than the value 
 provided by the Fermi's golden rule, but also different   from the standard expression.   Accordingly the revised probability   modifies a reflection rate of visible lights substantially.

 We compare  the new value with  direct observations in the ground level.   Reflected  light by  molecules in the atmosphere is symmetric in forward and backward directions. Using the scattering probability $P_R$, fraction  of  the direct sun light  and diffused light on the Earth's surface, and their ratio   $r_d$ are expressed by
\begin{eqnarray}
& &f_{direct}= 1-P_R , f_{diffuse}= \frac{P_R}{2},\\
& &r_d=\frac{f_{diffuse}}{f_{direct}}=\frac{P_R}{2(1-P_R)}. 
\end{eqnarray} 
Albedo is the  reflection rate of the sun light and is expressed with $r_d$ as
\begin{eqnarray}
\text {Albedo}=\frac{P_R}{2}=\frac{r_d}{1+2r_d}. \label{albedo}
\end{eqnarray}
The value is easily evaluated as 
 \begin{eqnarray}
 & & \text {Albedo}~~~~~0.25~~~~~0.27~~~~~0.29~~~~~0.30~~~~~0.32 \nonumber \\   ~~~
& &r_d ~~~~~~~~~~~~~  0.5~~~~~0.6~~~~~~0.7~~~~~~~0.8~~~~~~0.9
\end{eqnarray} 
Seasonal change of $r_d$ is large. Here we use a value at Sapporo,  $r_d=0.58$, \cite{ JMA_data}, then  we have  $ \text{ Albedo}=0.27$. This  agrees well with the latest value.

 {\bf Miscellaneous  problems (i) Lights of small wave packets }

For  small $\sigma_{\gamma} $,  the $A$ term is dominant and  the $B$ term is negligible in Eq.$(\ref{amplitude_s2}  )$.  The  probability  is in agreement   with original  Rayleigh scattering \cite{Rayleigh}. Using this cross section, $\sigma_{Rayleigh}$,  and moleculer density, $\rho_M$,  a mean free path of visible light in the atmosphere, $L_{mfp}$ is expressed by   
$\frac{1}{\sigma_{Rayleigh} \rho_M}. $
 Substituting cross section $\sigma_{Rayleigh}$ presented in Supplement  and density $\rho_M $ of one atmosphere, we find  the mean free path is around $20$ kM   \cite{Rayleigh_{exp} } in the ground level. This length  is consistent with  day-to-day observations.   A value  in  clean water with number  density  of water $\frac{ 6 \times 10^{23} \times 10^6}{18 }$ per Meter$^3$ and  measured value in \cite{Rayleigh_{SK}} shows that  the mean free path is consistent with the naive  estimations. These facts verify   estimations of the mean free path in the atmosphere with    lowest order calculations. Higher order corrections   were shown small  in \cite{higher_order}.   Space-time view in these situations  is given in Fig.1


{ (ii) \bf Mie scattering with couds }
 In the Troposphere,   cloud activities are present. Scattering with large scatterers such as water droplet is described by Mie scattering. The scattered light has small backward component and  does not expand with time. Accordingly,the scattered light  reveals   clear shape of scatterers, 
%
%
%
%
and  the backward scattering is provided by   Rayleigh scattering. Accordingly,   Eq.$(\ref{albedo} )$ is the total reflection probability.

{ (iii) \bf  Infrared lights emitted from the Earth } 

 The   temperature of the Earth surface is abount $300$ K and the	lights from  black body radiation correspond  to  infrared light.   This light  in  the ground surface has much shorter  coherence length than the direct solar light and the wave packet size  
around  micro meters.  Because this size  is much smaller  than the wave packet size of solar light, the second class quantities are small.  Absorptions  by molecules are present and 
have  much larger probability than  the  Rayleigh scattering  Eq.$(\ref{Probability_{Rayleigh}})$.

{\bf Laser nanoparticle Rayleigh  experiment.} 

In laboratoy experiments, the wave packet sizes  depend on experimental setup.  Incident  laser light has a  long coherence length. Its coherence lengths  are much larger than atomic sizes and     huge  in the longitudinal direction.   In the following we study  the wave packets  of the initial state  described by symmetric wave packet for simplicity. The result is applied to   asymmetric wave packet as well.  Scattered  light is  measured with photo-multiplyer, and the size of the final gamma is determined by the photon detector and satisfy  $\sigma_{\gamma_1} \gg \sigma_{\gamma_2} , \sigma_M$, 
and the spatial  and  temporal sizes are $
\sigma_s=\frac{\sigma_{\gamma_2} \sigma_M} { \sigma_M+2 \sigma_{\gamma_2}} 
$ and   $\frac{1}{2\sigma_t}=  \frac{1}{ (\sigma_M+2\sigma_{\gamma_2})} V_2^2$.  Other quantities are
${\vec V}_0=  \frac{ \sigma_M} { \sigma_M+2 \sigma_{\gamma_2}} {\vec V}_{\gamma_2}$
$\delta  \omega
 =V_{\gamma}P_{\gamma_1}(1-\frac{\sigma_M}{\sigma_M+2 \sigma_{\gamma_2}}\cos \theta ) -V_{\gamma}P_{\gamma_2}(\frac{2\sigma_{\gamma_2}}{\sigma_M+2\sigma_{\gamma_2}} )$.
Position dependent factors are  
$ T_{int}=   -\frac{ \vec V_2 (\vec{\tilde X}_{\gamma_2}-\vec X_M ) }{ {  V_{\gamma_2}}^2 }$ and 
$ \frac{R}{2} 
 =  \frac{1 }{ (\sigma_M+2\sigma_{\gamma_2})}{(\vec {\tilde X}_{\gamma_2}-\vec X_M )_T}^2$.   

Kinematical factor that depends on the energies  and momenta are,
\begin{eqnarray}
& &\frac{\sigma_t}{2} (\delta \omega)^2+\frac{\sigma_s}{2} (\delta \vec P)^2 \nonumber \\
& &=\frac{ \sigma_{\gamma_2}}{2} (P_{\gamma_2}-P_{\gamma_1})^2+ \frac{ \sigma_{\gamma_2}}{(\sigma_M+2 \sigma_{\gamma_2})}[  (\frac{(\sigma_M+2 \sigma_{\gamma_2}) }{2 \sigma_{\gamma_2}}) (1 -\cos \theta )^2  +  (1- \cos^2 \theta)]\frac{\sigma_M}{2} P_{\gamma_1}^2 
\end{eqnarray}

Ignoring an interference term between the zero-oth order term and the first order term, and the one between $A$ term and $B$ term from reasons mentioned in the appendix, we have a probability distribution  E.$( \ref{Probability_{Rayleigh}})$. $\sigma_s$ and $\sigma_t$ are small and  the probability reveals  specific  energy dependence.

 The power behavior in the large $\omega$ shows clear signal of the second term. The transition probability is dominated by the second term  in the bracket in the  energy region    
\begin{eqnarray}
& & \sigma_t (\delta \omega)^2 \gg \frac{(T_{int}-T_0)^2}{\sigma_t},
\end{eqnarray}
where  the  right-hand side satisfies
\begin{eqnarray}
 \frac{ \sigma_t}{2\pi}   \frac{1}{(T_{int}-T_0)^2+({\sigma_t}  \delta \omega )^2} =\frac{1}{2\pi}\frac{1}{ {\sigma_t}  (\delta \omega )^2} \gg e^{-\sigma_t(\delta \omega )^2}.
\end{eqnarray}
Figure 2 shows the experimental value  from Laser nanoparticle experiment. 

(2)  $\vec X_M$ and $\vec X_{\gamma_2}$ are determined by experimental setups.  $\sigma_t$ is small, and $e^{-\frac{(T_{int}-T_0)^2}{\sigma_t}}$ is sizable in $ |(T_{int}-T_0)| \leq \sqrt \sigma_t$. Accordingly, the total value of second contribution is given by a factor $\frac{\sqrt \sigma_s}{V_{\gamma}(T_1-T_0)} \approx 10^{-9}$ of the first contribution. This value is extremely small
, but is not impossible to measure.  
  for small $\sigma_M$ should be useful for testing the formula. For (1) $\sigma_{\gamma} \approx  \sigma_M$, the scattered light is observed only at an on-axis region, whereas for (2) $\sigma_{\gamma} > \sigma_M$ the scattered light is observed at an off-axis region.     
 The first   one should be realized with  normal lamps and the second one should be realized with  lasers.  

{\bf Refractive index and  energy density of diffuse lights} 

One of frist class quantities  of a light is   refractive index.  This has the short-range origin and the value is connected with the $A$ term, but with $B$ term.
 An  energy distribution of the diffuse lights  at a position $\vec X_{\gamma_2}$  from  the scattering with a molecule at ${\vec X}_M$ is expressed using    Eq.$(\ref{Probability_{Rayleigh}} )$. 
 Rayleigh scattering is symmetric in a forward and backward diections.  The distribution of lights  in  forward direction $(\vec n_{\gamma_1}, \vec n_{\gamma_2}) >0$ is the same as  the one  in  backward  direction $(\vec n_{\gamma_1}, \vec n_{\gamma_2}) <0$.


A scattering of the light with free electrons is   the Thomson scattering.   Because the free electron has a charge,   the coupling strength $\kappa= \frac{e^2}{2m}$ does not decreas at $E \rightarrow 0 $.  Because this coupling is constant at  low energy, the bulk term is constant at low energy region.   $P^{(d)}$ behaves differently and gives a sizable contribution in the low energy.The electron mass is $\frac{1}{2000}$ of the proton mass, and 
the mean free path is estimated as
$l_e=    10^{-5} \text{M} $       .
This is much longer  than the atom's mean free path and wave lengths of visible lights.   The Thomson scattering in the atmosphere occurs mainly  in the forward angle, and has different  behaviers  form    Rayleigh scattering.   

\begin{figure}
\begin{center}
\includegraphics[width=120mm]{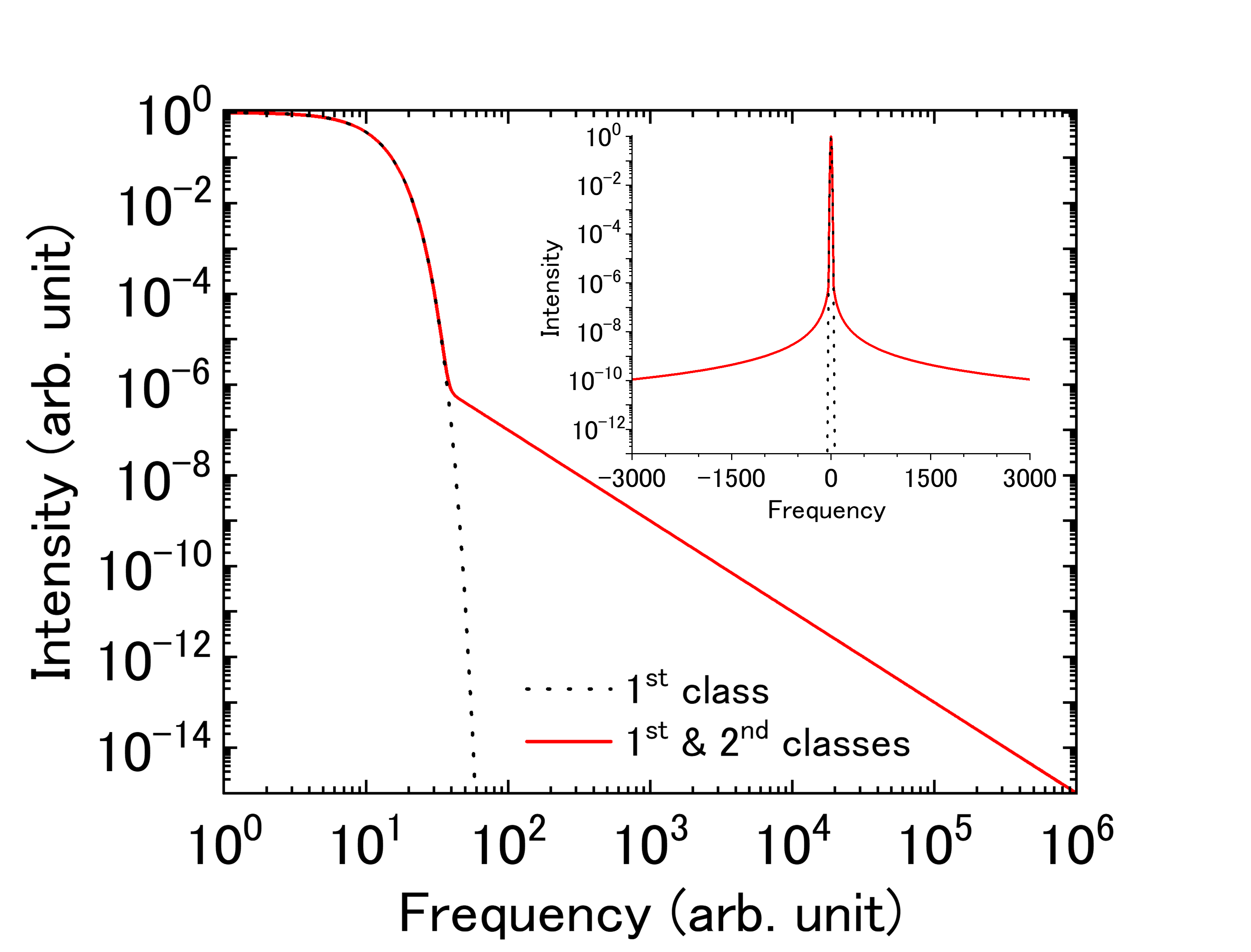} 
\caption{ Photon spectrum of Rayleigh scattering of laser light  with  nanoparticle. Holizontal line represents frequency shift and vertical line represent intensity. A sharp peak around  
the center is  the first class component and the broad tale decreasing with $\sim ( \Delta \nu)^{-2}$ represent the second class component.  
}
\end{center}
\end{figure}

{ \bf Summary and Implications}

The  study of light scattering  was initiatted  by Rayleigh \cite{Rayleigh}, and  the reason for the sky's blue color was clarified from the strength  proportional to $E^{4}$. Despite of qualitative agreement,  absolute  strength  of observed lights are  much higher  than the calculations. The devation is even  larger at high altitude of $10$ km suggesting that the probability  is not  proportional to the molecule density. Rayleigh scattering is only process in this energy range, and the disagreement is  a fundamental problem. In this paper, we have derived the new formula of the scattering probability, which  is composed of two terms, 
the first and second  classes as $P(T)=\Gamma T+P^{(d)}$ for $P(T) \ll 1$. Here  $\Gamma$ is the first class and $P^{(d)}$ is the second class. Contraly to the first class quantities, the second class quantities are absent in the previous expressions, and violate  the  kinetic-energy conservation and  scattering into cones theorem. Moreover these 
 depend on the wave-packet sizes.  For large wave-packet sizes, Rayleigh scattering  is very different from the classical one as seen in Fig.1.  Solar lights have large wave packt size, and their  Rayleigh scattering has the observed strengh and properties. Our theoretical value of Albedo  and the strength of scattered  light at the height $10$ km agree with observations.  Rayleigh scattering  for the short wave packets is similar to the classical one, and  has  the extremely small  second class quantity, as seen in laboratory experiments in 
  Fig.2. The present theory explains the color of the sky and the laser experiment.   


\subsection*{Acknowledgments}
 This work was partially supported by a 
Grant-in-Aid for Scientific Research ( Grant Number 24340043 and Grant Number 26K00623).



\appendix 
\section{  Wave packet formalism for Rayleigh scattering }

In the main text, we present the new  formula of  Rayleigh scattering of solar lights with molecules in the atmosphere, and the one of laser lights with nanopartilce. The detailed derivations are presented in Appendix.




\subsubsection{ Light molecule  scattering}


Interaction Hamiltonian of a photon with a particle of an  electric dipole moment is
\begin{eqnarray}
H_{int}=g \int d {\vec x}F_{0 i}(x)^2 \phi_M(x)^2,
\end{eqnarray}
where $\phi_M(x)$ represnts  the particle field.  Rayleigh scattering of an atom is described by an equivalent expressesion \cite{Rayleigh_{exp} }.

A Gaussian wave packet in the coordinate representation for a particle of the energy $E(\vec p)$ and a group velocity  $\vec v_0$ is 
\begin{eqnarray}
\langle t,{\vec x}| {\vec P}_0, {\vec X}_0, T_0  \rangle&=& (\sigma \pi )^{-3/4} e^{-\frac{1}{2 \sigma }( \vec x-{\vec X}_0 -{\vec v}_0(t-T_0))^2 -i\frac{E({\vec P}_0)}{\hbar}(t-T_0)+i\frac{{\vec P}_0}{\hbar}({\vec x}-{\vec X}_0)}   \label{wave-packet_3}, \nonumber \\
 v_0^i &=& \frac{\partial}{\partial p_i}E(p)|p=P_0,
\end{eqnarray}
where $\vec P_0$ is a central momentum and $\vec X_0$ is a central position at $t=T_0$.   The wave function  is approximately  plane wave  in the spatial region $\frac{|{\vec x}-{\vec X}_0|}{  \sqrt \sigma} \leq 1$ and vanishes at large  $\frac{|{\vec x}-{\vec X}_0|}{\sqrt \sigma} \gg 1$ at $t=T_0$. At a finite  time the wave packet moves, while  expansion is negligible,  and the central position becomes ${\vec X}_0 +{\vec v}_0(t-T_0) $. At an extremely larger time, the expansion is not negligible.  This paper studies the intermediate situation, which is rellevant to standard experiments.   Wave packets are normalized $ \langle \vec P, \vec X| \vec P,\vec X \rangle =1$ and satisfy the completeness. 
Any  normalized vector is expresse by 
$ \int \frac{d^3 p d^3 X}{(2\pi)^3 } |{\vec P}, {\vec  X} \rangle \langle {\vec P}, {\vec  X} |\Psi \rangle =|\Psi \rangle$ with the norm
$\int \frac{d^3 p d^3 X}{(2\pi)^3 }  |\langle {\vec P}, {\vec  X} |\Psi \rangle|^2 =\langle \Psi |\Psi \rangle$. In the  Gaussian wave packets,
the matrix element of the molecule  field of the wave packet of  the size $\sigma_{M}$ in the energy $E(\vec P_M )$, the momentum     $\vec P_M $, and the position $\vec X$, 
and the  photon,
\begin{eqnarray}
\langle  0 |\varphi_M(x)|{\vec P}_M, {\vec  X}_M \rangle
	&=&	N_M  \eta_M  \left(\frac{2\pi}{\sigma_M }\right)^{3/2}  e^{-\frac{1}{2 \sigma_M}  ({\vec x}-{\vec X}_M-{\vec V}_M(t-T_M))^2 -i(\tilde E({\vec P}_M) (t-T_M )-{\vec P}_M\cdot({\vec x}-{\vec X}_M )) },  
  \label{M_wave packet1} \nonumber \\
\langle {\vec P_{\gamma_i}},{X}_i|A^{\mu}(x)|0 \rangle
	&=&	N_i \eta_{\gamma_i} \left(\frac{2\pi}{\sigma_{\gamma_i}}\right)^{3/2}
		\epsilon^{\mu} (P_{\gamma_i}) \times
e^{-\frac{1}{2 \sigma_{\gamma_i}}  ({\vec x}-{\vec X}_{\gamma_i}-{\vec V}_{\gamma_i} ({t-T_{\gamma_i}}))^2} \nonumber  \\
& &e^{+i({E({\vec P}_{\gamma_i}) ({t-T_{\gamma_i}})-{\vec P}_{\gamma_i} \cdot ({{\vec x}-{\vec X}_{\gamma_i}})})}, \label{approximate wave packet_{photon}}
\end{eqnarray}
with $
{\vec V}_M=\frac{{\vec P}_M}{E({\vec P}_M)}, E({\vec P}_M)=\sqrt{{{\vec P}_M}^2+M_M^2 } ,  N_M= (\frac{\sigma_M}{\pi})^{3/4}, \eta_M =(\frac{1}{(2\pi)^3 2E_M})^{1/2} $
and variables for  photon, and  for a molecule
\begin{eqnarray}
& &\langle  0 |\varphi_M(x)|0, {\vec  X}_M \rangle
	=	N_M  \eta_M  \left(\frac{2\pi}{\sigma_M }\right)^{3/2}  e^{-\frac{1}{2 \sigma_M}  ({\vec x}-{\vec X}_M)^2-i M_M (t-T_M )  }.
  \label{M_wave packet2}  
\end{eqnarray}
 In a limit $M \rightarrow \infty$,  the initial molecule and final molecule are at rest and, ${\vec P}_M={\vec V}_M=0$ and $E_M=M_M$
The molecule is at rest of the momentum and position  $ \vec X_M=0, \vec P_M=0 $. Wave packets are summarized in ref. \cite{ishikawa-shimomura-PTP,ishikawa-jinnouchi,Ishikawa-Nishio, Ishikawa_1,Ishikawa-Tobita-ptp,Ishikawa-Tobita_1}. 
\subsubsection{Transition amplitude}


Strength of an initial gamma  is
$( \sigma_{\gamma})^{-3/4} \frac{1}{\sqrt{ 2E_{\gamma_1}}}$.
The relative strength of a final state is obtained by dividing  the matrix element by this factor.

The transition amplitude in the intreraction picture is given by Dyson formula,   for a state specified by $\alpha$ and $\beta$, $\langle \Psi_{{\text i},\beta}| U(T,0)|\Psi_{i,\alpha} \rangle$, 
\begin{eqnarray}
& &U(T,0)=1+g \int_0^T dt_1 \frac{H_{{\text i},int}(t_1)}{i\hbar}+g^2\int_0^T dt_1\int_0^{t_1} dt_2 \frac{H_{{\text i},int}(t_1)}{i\hbar}\frac{H_{{\text i},int}(t_2)}{ i\hbar} +\cdots, 
\end{eqnarray}

The lowest oder amplitude $S_0$ is a matrix element of the wave packets and is given by 
\begin{eqnarray}
& &S_0=\langle {\vec P}_{\gamma_1 }, {\vec  X}_{\gamma_1 },T_1| {\vec P}_{\gamma_2 }, {\vec  X}_{\gamma_2 },T_2 \rangle= e^{-\frac{1}{4 \sigma_{\gamma}}({\vec X}_{\gamma_1 }-{\vec X}_{\gamma_2 } -{\vec V}_0 (T_1-T_2))^2-\frac{\sigma_{\gamma}}{4} ({\delta \vec P})^2 +i \theta_0} \label{zeroth_am},\nonumber \\ 
& & {\vec V}_0=\sigma_s( \frac{{\vec V}_{\gamma_1}}{\sigma_{\gamma_1}} +\frac{{\vec V}_{\gamma_2 }}{\sigma_{\gamma_2}}), ~~ \delta {\vec P}=  {\vec P}_{\gamma_1}-{\vec P}_{\gamma_2}  ,~i \theta_0=-i(E_{\gamma_1}T_{\gamma_1}-\vec p_{\gamma_1} \vec X_{\gamma_1}-E_{\gamma_2}T_{\gamma_2}+\vec p_{\gamma_2} \vec X_{\gamma_2}) \nonumber \\
& & 
\end{eqnarray}

The first  oder amplitude $S_1$ is obtained by substituting Eqs.$(  \ref{M_wave packet1}  )$ and $ (\ref{M_wave packet2})$   to the S-matrix element, and integrating  over the entire space, $-\infty \leq x_i \leq \infty$, and $0\leq t \leq T$. Then we have  the amplitude
\begin{eqnarray}
S_1&=&i  \kappa \left(\prod_{A=1}^4(\pi \sigma_A)^{-3/4} \frac{1}{\sqrt 2E_A} \right)  V(P_i,T_{int}), \label{amplitude_s}   \nonumber \\
 & &V(P_i,T_{int})= \int_{T_0}^{T_1}  dt  \int d \vec x e^{-f(\vec x)},   \nonumber \\
& &f(\vec x)=\frac{1}{2 \sigma_{\gamma_1}}(\vec x-\vec V_{\gamma_1} t -\vec    {\tilde   X}_{\gamma_1})^2+\frac{1}{2 \sigma_{\gamma_2}}(\vec x-\vec V_{\gamma_2} t -\vec {\tilde X}_{\gamma_2})^2+\frac{2}{2 \sigma_M}(\vec x- \vec X_M)^2  \nonumber \\
& &~+iE_{\gamma_1}(t-T_1)-iE_{\gamma_2}(t-T_2)-i\vec P_{\gamma_1} (\vec x-\vec X_{\gamma_1})+i \vec P_{\gamma_2} (\vec x-\vec X_{\gamma_2}).  
\end{eqnarray}

 Using  solution of  $f'(\vec x_0)=0$, 
\begin{eqnarray}
& &\vec x_0=\sigma_s ((\frac{1}{ \sigma_1}\vec V_{\gamma_1} +\frac{1}{ \sigma_{\gamma_2}}\vec V_{\gamma_2} )t+(\frac{1}{ \sigma_{\gamma_1}}\vec {\tilde X}_{\gamma_1}+\frac{1}{ \sigma_{\gamma_2}}\vec {\tilde  X}_{\gamma_2}+\frac{2}{ \sigma_M}\vec X_M)+  i(\vec P_{\gamma_1}-\vec P_{\gamma_2})), \nonumber \\
& &\frac{1}{\sigma_s}=\frac{1}{\sigma_{\gamma_1}}+\frac{1}{\sigma_{\gamma_2}}+\frac{2}{\sigma_M}=\frac{D}{\sigma_1 \sigma_2 \sigma_M}, \nonumber \\
& &\sigma_s=\frac{\sigma_{\gamma_1} \sigma_{\gamma_2} \sigma_M} {D},~~D= {(\sigma_{\gamma_1} +\sigma_{\gamma_2}) \sigma_M+2\sigma_{\gamma_1} \sigma_{\gamma_2}}. \label{parameter1}
\end{eqnarray}
The exponent is expressed as  
$f( \vec x)=f(\vec x_0)+\frac{1}{2\sigma_s }(\vec x -\vec x_0)^2$,
$f(\vec x_0)=C_2 t^2+2C_1 t +C_0$,
where
\begin{eqnarray}
& &C_2=\frac{1}{2 \sigma_1}(\frac{\sigma_s-\sigma_{\gamma_1}}{ \sigma_{\gamma_1}}\vec V_{\gamma_1} +\frac{\sigma_s}{ \sigma_{\gamma_2}}\vec V_{\gamma_2} )^2+\frac{1}{2 \sigma_{\gamma_2}}(\frac{\sigma_s}{ \sigma_1}\vec V_{\gamma_1} +\frac{\sigma_s-\sigma_{\gamma_2}}{ \sigma_{\gamma_2}}\vec V_{\gamma_2} )^2+{\bf 2}\frac{1}{2 \sigma_M}(\frac{\sigma_s}{ \sigma_1}\vec V_{\gamma_1} +\frac{\sigma_s}{ \sigma_{\gamma_2}}\vec V_{\gamma_2} )^2 \nonumber\\ 
& & \label{coeffieints_1}\\
 & &C_1=\frac{1}{2 \sigma_{\gamma_1}}(\frac{\sigma_s-\sigma_{\gamma_1}}{ \sigma_{\gamma_1}}\vec V_{\gamma_1} +\frac{\sigma_s}{ \sigma_{\gamma_2}}\vec V_{\gamma_2} )[ \frac{\sigma_s-\sigma_{\gamma_1}}{ \sigma_{\gamma_1}}\vec{\tilde X}_1+\frac{\sigma_s}{ \sigma_{\gamma_2}}\vec{\tilde X}_{\gamma_2}+\frac{2\sigma_s}{ \sigma_M}\vec X_M ]\nonumber \\
& &+\frac{1}{2 \sigma_2}(\frac{\sigma_s}{ \sigma_{\gamma_1}}\vec V_{\gamma_1} +\frac{\sigma_s-\sigma_{\gamma_2}}{ \sigma_2}\vec V_{\gamma_2} )[\frac{\sigma_s}{ \sigma_{\gamma_1}}\vec{\tilde X}_{\gamma_1}+\frac{\sigma_s-\sigma_{\gamma_2}}{ \sigma_{\gamma_2}}\vec{\tilde X}_{\gamma_2}+\frac{2\sigma_s}{ \sigma_M}\vec X_M ]\nonumber \\
& &+{\bf 2}\frac{1}{2 \sigma_M}(\frac{\sigma_s}{ \sigma_{\gamma_1}}\vec V_{\gamma_1} +\frac{\sigma_s}{ \sigma_{\gamma_2}}\vec V_{\gamma_2} )[\frac{\sigma_s}{ \sigma_{\gamma_1}}\vec{\tilde X}_{\gamma_1}+\frac{\sigma_s}{ \sigma_{\gamma_2}}\vec{\tilde X}_{\gamma_2}+\frac{2\sigma_s-\sigma_M}{ \sigma_M}\vec X_M ]  +i\frac{\delta \omega }{2}, 
\nonumber \\
 \label{coeffieints_2} 
& &C_0=\frac{1}{2 \sigma_{\gamma_1}}[ \frac{\sigma_s-\sigma_{\gamma_1}}{ \sigma_1}\vec{\tilde X}_{\gamma_1}+\frac{\sigma_s}{ \sigma_{\gamma_2}}\vec{\tilde X}_{\gamma_2}+\frac{2\sigma_s}{ \sigma_M}\vec X_M + i\sigma_s  (\vec P_{\gamma_1}-\vec P_{\gamma_2})]^2\nonumber \\
& &+\frac{1}{2 \sigma_2}[\frac{\sigma_s}{ \sigma_{\gamma_1}}\vec{\tilde X}_{\gamma_1}+\frac{\sigma_s-\sigma_{\gamma_2}}{ \sigma_2}\vec{\tilde X}_{\gamma_2}+\frac{2\sigma_s}{ \sigma_M}\vec X_M + i\sigma_s  (\vec P_{\gamma_1}-\vec P_{\gamma_2})]^2\nonumber \\
& &+{\bf 2}\frac{1}{2 \sigma_M}[\frac{\sigma_s}{ \sigma_{\gamma_1}}\vec{\tilde X}_{\gamma_1}+\frac{\sigma_s}{ \sigma_{\gamma_2}}\vec{\tilde X}_{\gamma_2}+\frac{2\sigma_s-\sigma_M}{ \sigma_M}\vec X_M+ i\sigma_s  (\vec P_{\gamma_1}-\vec P_{\gamma_2})]^2 ~-iE_{\gamma_1} T_1+iE_{\gamma_2} T_2 \nonumber \\
& &~+i\vec P_{\gamma_1} \vec X_{\gamma_1} -i \vec P_{\gamma_2}  \vec X_{\gamma_2}-i(\vec P_{\gamma_1}-\vec P_{\gamma_2})[ \frac{\sigma_s}{ \sigma_{\gamma_1}}\vec{\tilde X}_{\gamma_1}+\frac{\sigma_s}{ \sigma_{\gamma_2}}\vec {\tilde X}_{\gamma_2}+\frac{2\sigma_s}{ \sigma_M}\vec X_M +i\sigma_s  (\vec P_{\gamma_1}-\vec P_{\gamma_2})].  \label{coeffieints_3}
\end{eqnarray}
In the above equation,   $\delta \omega $ is the angular velocity, which is  the  energy of overlaping waves and depends upon the momentum difference $\delta {\vec P}$ and the velocities, 
\begin{eqnarray} 
& &\delta  \omega=  \delta E-{\vec V}_0\cdot {\delta \vec P},~~{\delta E}=E_{\gamma_1}-E_{ \gamma_2},~~
 \delta {\vec P}=  {\vec P}_{\gamma_1}-{\vec P}_{\gamma_2}  \nonumber \\
& &{\vec V}_0=\sigma_s( \frac{{\vec V}_{\gamma_1}}{\sigma_{\gamma_1}} +\frac{{\vec V}_{\gamma_2 }}{\sigma_{\gamma_2}})\label{parameter2} .  
\end{eqnarray}
 $f(\vec x_0)$  is  quadratic in time around the center $t_{0}$,$f(\vec x_0)=  +C_2(t-t_{0})^2+C_0-\frac{C_1^2}{C_2}, $
where 
\begin{eqnarray}
t_{0}=-\frac{C_1}{C_2}=T_{int}-i\sigma_t \delta \omega, 
\end{eqnarray}
where real part $T_{int}$ represent the intersection time of  the initial and final states.  

(1)  For the solar lights in the atmosphere,  $\sigma_{\gamma_1}=\sigma_{\gamma_2} =\sigma_{\gamma} \gg \sigma_{M}$ and $\sigma_s=\frac{\sigma_M}{2}  \ll \sigma_{\gamma}$,
\begin{eqnarray}
& &C_2=\frac{1}{ \sigma_{\gamma}}(\vec V_{\gamma_1}  )^2 \nonumber\\ 
 & &C_1=\frac{1}{2 \sigma_{\gamma}}\vec V_{\gamma_1}  \vec{\tilde X}_1+\frac{1}{2 \sigma_{\gamma}}    \vec V_{\gamma_2}   \vec{\tilde X}_{\gamma_2}  +i\frac{\delta \omega }{2} \nonumber \\ 
& &C_0=\frac{1}{2 \sigma_{\gamma}}[ -\vec{\tilde X}_{\gamma_1}+ \vec X_M]^2 +\frac{1}{2 \sigma_{\gamma}}[-\vec{\tilde X}_{\gamma_2}+\vec X_M ]^2+ \frac{\sigma_M}{4}[ (\vec P_{\gamma_1}-\vec P_{\gamma_2})]^2  \nonumber \\
& &~-iE_{\gamma_1} T_1+iE_{\gamma_2} T_2+i\vec P_{\gamma_1} \vec X_{\gamma_1} -i \vec P_{\gamma_2}  \vec X_{\gamma_2}
-i(\vec P_{\gamma_1}-\vec P_{\gamma_2}) \vec X_M. 
\end{eqnarray}
In  spatial  region $ | \vec{\tilde X}_{i} | \ll \sqrt \sigma $, the exponent is independent of positions,
\begin{eqnarray}
& &C_0-\frac{C_1^2}{C_2}  =\frac{\sigma_M}{4} (\vec P_{\gamma_1}-\vec P_{\gamma_2})^2 +\frac{\sigma_{\gamma}}{4}(\delta \omega)^2 -iE_{\gamma_1} T_1+iE_{\gamma_2} T_2+i\vec P_{\gamma_1} \vec X_{\gamma_1} -i \vec P_{\gamma_2}  \vec X_{\gamma_2}
-i(\vec P_{\gamma_1}-\vec P_{\gamma_2}) \vec X_M. \nonumber  \\
\end{eqnarray}

(2)  In laser experiments in laboratory, $\sigma_{\gamma_1} \gg \sigma_{\gamma_2} \sim  \sigma_{M}$, 
$\sigma_s=\frac{\sigma_{\gamma_2} \sigma_M} {\sigma_M+2 \sigma_{\gamma_2}},~~D= \sigma_{\gamma_1}( \sigma_{M}+2\sigma_{\gamma_2})$, 
\begin{eqnarray}
& &C_2= \frac{1  }{ \sigma_M+2 \sigma_{\gamma_2}}(\vec V_{\gamma_2} )^2   \label{coeffieints_1}, \nonumber \\
 & &C_1= \frac{1} {\sigma_M+2 \sigma_{\gamma_2}} \vec V_{\gamma_2} [ \vec{\tilde X}_{\gamma_2}-\vec  X_M ] +i\frac{\delta \omega }{2},\nonumber \\
& &C_0 =\frac{1}{\sigma_M+2 \sigma_{\gamma_2}} (\vec{\tilde X}_{\gamma_2}-\vec X_M)^2  +\frac{\sigma_s}{2} (\vec P_{\gamma_1}-\vec P_{\gamma_2})^2 -i(\vec P_{\gamma_1}-\vec P_{\gamma_2} )[+\frac{ \sigma_M} {\sigma_M+2 \sigma_{\gamma_2}}  \vec {\tilde X}_{\gamma_2}+\frac{ 2 \sigma_{\gamma_2}} {\sigma_M+2 \sigma_{\gamma_2}} \vec X_M ]\nonumber \\
 & & -iE_{\gamma_1} T_1+iE_{\gamma_2} T_2+i\vec P_{\gamma_1} \vec X_{\gamma_1} -i \vec P_{\gamma_2}  \vec X_{\gamma_2}.
 \end{eqnarray}
 The vertex is summarizes as
 \begin{eqnarray} 
 & &V(P_i,T_{0})
=  (2 \sigma_s \pi )^{3/2}  e^{-( C_0-\frac{C_1^2}{C_2})}  \int_{T_0}^{T_1}  dt   e^{-C_2(t-t_{0}  )^2}  \nonumber \\
& &= (2 \pi \sigma_s)^{3/2}(2\pi \sigma_t)^{1/2}e^{-\frac{\sigma_s}{2}(\delta P)^2-R/2+i\phi_0}  G(T_{int}),
 \end{eqnarray} 
 with
$\frac{1}{\sigma_t}=2C_2$,$ T_{int}=-\frac{C_1}{C_2} +i \sigma_t \delta \omega$  
$\frac{R}{2} =C_0-\frac{C_1^2}{C_2}-\frac{\sigma_s}{2}(\delta \vec p)^2 $
and   
\begin{eqnarray}
 G(T_{int})&=&\int_{T
_0}^{T_1} dt   [{ \frac{1}{\sqrt{ 2 \pi  \sigma_t}}e^{-\frac{1}{2 \sigma_t} (t-T_{int}+ \sigma_t (i \delta \omega))^2+   \frac{\sigma_t}{2} (i \delta \omega )^2}} ]
 = A+B, \label{time_amplitude}\\
 A&=& \frac{1}{2} e^{\frac{\sigma_t}{2} (i \delta \omega )^2 } \left[sgn(T_{int}-T_0 )- sgn(T_{int}-T_1) \right], \nonumber \\
B&=&- \frac{1}{2}     e^{-\frac{(T_{int}-T_0)^2}{2 \sigma_t} +i(\delta  \omega )(T_{int}-T_0)} \sqrt{\frac{2 \sigma_t}{\pi}}  \frac{1}{T_{int}-T_0 - i\sigma_t (\delta  \omega )} \nonumber \\
 &+& \frac{1}{2}   e^{-\frac{(T_{int}-T_1)^2}{2 \sigma_t} +i(\delta   \omega)   (T_{int}-T_1)} \sqrt{\frac{2 \sigma_t}{\pi}} \frac{1}{T_{int}-T_1- i\sigma_t (\delta  \omega ) }. \nonumber 
\end{eqnarray}
 At large $T_1$, the last term in the first line becomes $-1/2$,  The final form of transition  amplitude  in the first order of the interaction,  
\begin{eqnarray}
S_1&=&i  \kappa \left(\prod_{A=1}^4(\pi \sigma_A)^{-3/4} \frac{1}{\sqrt{ 2E_A}} \right) e^{-\frac{\sigma_s}{2}( \delta \vec P)^2-\frac{R}{2} +i\theta_0}  (2 \pi \sigma_s)^{3/2}(2 \pi \sigma_t)^{1/2} (A+B)  , \label{amplitude_s}   \nonumber \\
& &\kappa=g(p_{\gamma_1}^0\epsilon^i(p_{\gamma_1})-p_{\gamma_1}^i{\epsilon(p_{\gamma_1})}^0( { p_{\gamma_2}}_0\epsilon_i({p_{\gamma_2}})-{p_{\gamma_2}}_i\epsilon_0(p_{\gamma_2})), \label{first_am}
\end{eqnarray}

The  amplitude and the transition probability are finite.  $G(T_{int})$ is a specific factor expressed by the error function, the details of which were  given  in  \cite{ Ishikawa-Oda} and in the Appendix.

Previous applications are found in \cite{Ishikawa-Tobita_1, Ishikawa-Tobita_2,  Ishikawa-Tajima-Tobita, Ishikawa-Oda, Ishikawa-Nishiwaki-Oda, Ishikawa-Nishiwaki-Oda_2, Ishikawa-Nishiwaki-Jinnouchi-Oda_2} 

\subsubsection{  Evaluation of the probability   } 
 The probability is evaluated for the case $\delta \omega=V_{\gamma}(P_2-P_1)$.
We study the S-matrix up to the first order in the coupling strength, $S_0+S_1$, Eqs.$(\ref{zeroth_am})$ and $(\ref{first_am})$
using Eq.$( \ref{time_amplitude})$. The probability is expressed as   
\begin{eqnarray}
|S|^2=|S_0|^2+S_0S_1^{*} +S_1S_0^{*}+|S_1|^2,
\end{eqnarray}
where  
\begin{eqnarray}
  |S_0|^2 &=&  e^{-\frac{1}{2 \sigma_{\gamma}}({\vec X}_{\gamma_1 }-{\vec X}_{\gamma_2 } -{\vec V}_0 (T_1-T_2))^2-\frac{\sigma_{\gamma}}{2} ({\delta \vec P})^2}, \nonumber \\
   S_0 S_1^{*} +S_1 S_0^{*} & =& e^{-\frac{1}{4 \sigma_{\gamma}}({\vec X}_{\gamma_1 }-{\vec X}_{\gamma_2 } -{\vec V}_0 (T_1-T_2))^2-\frac{\sigma_{\gamma}}{4} ({\delta \vec P})^2 -\frac{\sigma_t}{2}(\delta \bar \omega)^2-\frac{\sigma_s}{2}( \delta \vec P)^2-\frac{R}{2} } \nonumber   \\
  & \times&  \tilde N  (A+B) + h. c. 
\end{eqnarray}
Hereafter   we study a case of large $\sigma$  and small $\sigma_M$. Then $|\delta \vec p| $ is limited to  a small value. $|S_0|^2$ 
and $|S_0 {S_1}^{*}|$ are finite only in  the extremely  forward direction, and are presented  in a next paper.     
 
 The  probability from the first order amplitude is 
\begin{eqnarray}
& &|S_1|^2=  \kappa^2 \left(\prod_{A=1}^4(\pi \sigma_A)^{-3/4} \frac{1}{\sqrt{ 2E_A}} \right)^2  (2 \pi \sigma_s)^{3} (2 \pi \sigma_t) \times e^{-{\sigma_s}( \delta \vec P)^2-{R}   } J,    \\
& &J=(|A|^2 + BA^{*}+AB^{*}+ |B|^2 ), \nonumber 
\end{eqnarray}
where
\begin{eqnarray}
& &\left(\prod_{A=1}^4(\pi \sigma_A)^{-3/4} \frac{1}{\sqrt{ 2E_A}} \right)^2 =(\pi \sigma_{\gamma_1})^{-3/2}(\pi \sigma_{\gamma_2})^{-3/2}(\pi \sigma_{M})^{-3}\frac{1}{ 2E_{\gamma_1} 2E_{\gamma_2} (2M)^2},
\end{eqnarray}
and is summarizd as Eq,$(5)$ in the main text.

(1) Integration over the positions with a fixed momentum. 

Integration over the position is made with the longitudinal and tranaverse positions 
$\int d^3 X_{\gamma_2}= \int d {X_{\gamma_2}}_L  d^2 {X_{\gamma_2}}_T$.
The integration over transversal position is easily  made with the Gaussian integrals, and  the longitudinal position is converted to the intersection time, $T_{int}$. 

(1-1) The bulk term: $|A|^2$ for  $\omega=V_{\gamma}(P_{\gamma_2}-P_{\gamma_1})$
\begin{eqnarray}
2 \pi \int dX_{\gamma_2,L}P_{\gamma_2} dP_{\gamma_2} |A|^2&=&2 \pi\frac{P_{\gamma_1}}{V_{\gamma}} \int d \delta \omega e^{\sigma_t (-(\delta \omega)^2} \int_{T_0}^{\infty} d T_{int}   \left[sgn(T_{int}-T_0 )-sgn(T_{int}-T_1) \right]\nonumber\\
 &=& 2 \pi\frac{P_{\gamma_1}}{V_{\gamma}}(T_1-T_0)  \frac{\sqrt \pi}{\sqrt \sigma_t} 
\end{eqnarray}
The time is factorized from the momentum.

(1-2) The  boundary  term:  $|B|^2$:

(1-2-1) In laser laboratory experiments, $ \sqrt \sigma_t << V_{\gamma}(T_1-T_0)$, and $ e^{  -\frac{1}{\sigma_t}(T_{int}-T_0)^2   } $ decreases rapidly.
Using the variable $\xi=\sqrt{\frac{1}{\sigma_t}}(T_{int}-T_0) $,
\begin{eqnarray}
 \int_{T_0}^{T_1} dT_{int}  e^{  -\frac{1}{\sigma_t}(T_{int}-T_0)^2   } =\sqrt{\sigma_t}  \int _{0}^{\infty} d \xi e^{-\xi^2} =\sqrt{\sigma_t} \frac{\sqrt {\pi}}{2},  
\end{eqnarray}
the integrals of the boundary  term,   $|B|^2$,  is computed as,
\begin{eqnarray}
2 \pi \int dX_{\gamma_2,L}P_{\gamma_2} dP_{\gamma_2} | B|^2&= &2 \pi(\frac{\sqrt{2\sigma_t}}{2\sqrt \pi})^2\frac{P_{\gamma_1}}{V_{\gamma}} \int d \delta \omega \int_{T_0}^{\infty} dT_{int}   e^{  -\frac{1}{\sigma_t}(T_{int}-T_0)^2   } \frac{1}{\sigma_t^2(\delta \bar \omega )^2+(T_{int}-T_0)^2} \nonumber \\
 & &  =2 \pi\frac{1}{4\sqrt{ \pi \sigma_t}} \frac{P_{\gamma_1}}{V_{\gamma}} \int d \delta \omega \frac{1}{{\delta \bar \omega}^2+\frac{1}{2\sigma_t}} =2 \pi\frac{\sqrt {2 \pi}}{4} \frac{P_{\gamma_1}}{V_{\gamma}}.
\end{eqnarray}
The ratio of the boundary term over the bulk term is
\begin{eqnarray}
\frac{\int dX_{\gamma_2,L}  dP_{\gamma_2} | B|^2}{\int d X_{\gamma_2}P_{\gamma_2} dP_2 | A|^2}=\frac{\frac{\sqrt {2 \pi}}{4} \frac{P_{\gamma_1}}{V_{\gamma}} }{ \frac{P_{\gamma_1}}{V_{\gamma}}(T_1-T_0)  \frac{\sqrt \pi}{\sqrt  \sigma_t}  }=
\frac{ \sqrt{2 \sigma_t}}{4(T_1-T_0) } \ll 1 (\text{Laboratory})
\end{eqnarray}
In experiments and natural phenomena of small $\sigma_t$, the boundary terms are  much smaller than the bulk term generally. The value is $10^{-9}$ in Fig.2.  

(1-2-2) In solar light scatterings, $ \sqrt \sigma_t >> V_{\gamma}(T_1-T_0)$.
In  integral over  $T_{int}-T_0 $,  $ e^{  -\frac{1}{\sigma_t}(T_{int}-T_0)^2   } \sim 1$.
The integrals of the boundary  term,  $|B|^2$,  is written as,
\begin{eqnarray}
2 \pi\int dX_{\gamma_2}P_{\gamma_2} dP_{\gamma_2} |B|^2&= &2 \pi(\frac{\sqrt{2\sigma_t}}{2\sqrt \pi})^2 \frac{P_{\gamma_1}}{V_{\gamma}}\int d \delta \omega \int_{T_0}^{T_1} dT_{int}   e^{  -\frac{1}{\sigma_t}(T_{int}-T_0)^2   } \frac{1}{\sigma_t^2(\delta \bar \omega )^2+(T_{int}-T_0)^2}  \nonumber\\
& & =2 \pi\frac{P_1}{V_{\gamma}}\int d \delta \omega \frac{1}{4 \delta \omega}=2 \pi\frac{P_1}{V_{\gamma}}\frac{1}{4} \log \frac{\omega_{max}}{\omega_{mini}}, 
 \end{eqnarray}
 where $\omega_{max}$ and $\omega_{min}$ are cut off parameters in ultraviolet and infrared regions. 
This is independent of $\sigma_t$.    
The ratio of the boundary term over the bulk term is
\begin{eqnarray}
\frac{\int d X_{\gamma_2,L} P_{\gamma_2} dP_{\gamma_2} | B|^2}{\int dX_{\gamma_2}P_{\gamma_2} dP_{\gamma_2} | A|^2}=\frac{\frac{\log{\frac{\omega_{max}}{\omega_{min}}}}{4}  \frac{P_{\gamma_1}}{V_{\gamma}} }{ \frac{P_{\gamma_1}}{V_{\gamma}}(T_1-T_0)  \frac{\sqrt \pi}{\sqrt  \sigma_t}  }=
\frac{ \sqrt{ \sigma_t}}{4(T_1-T_0) } \log{\frac{\omega_{max}}{\omega_{min}}} \gg 1 (\text{Atmosphere})
\end{eqnarray}
and  causes  an enhancemnt in the probability. In experiments and natural phenomena of large  $\sigma_t$, the boundary terms are  much larger than those of small $\sigma_t$, and even larger  than the bulk term ocasionary. In these transitions, the boundary terms  have sizable  magnitudes  and  play crucial  roles.  

(1-3) The  integrals  of the cross term $A^{*}B$ (bulk-boundary interference term)
\begin{eqnarray}
& &\int_{T_0}^{\infty} d T_{int}   \frac{1}{2}   e^{-\frac{(T_{int}-T_0)^2}{2 \sigma_t}} 
\sqrt{\frac{2 \sigma_t}{\pi}}  \frac{1}{T_{int}-T_0+ i\sigma_t (\delta  \omega)}
=\sqrt{\frac{1}{2}}\frac{1}{ i \omega +\sqrt{2 \sigma_t}}. 
\end{eqnarray}

(2) Integrattion over the final states, in which  the momentum is integrated first and positions next, using
$\int \frac{d^3 P_{\gamma_2}}{2E_{\gamma_2}}= 2 \pi \int P_{\gamma_2} dP_{\gamma_2}$. 

(2-1) The bulk term, $|A|^2$. 
\begin{eqnarray}
\int dX_{\gamma_2,L}\frac{d^3 P_{\gamma_2}}{2E_{\gamma_2}} |A|^2&=&\int \frac{d^3 P_{\gamma_2}}{2E_{\gamma_2}} e^{\sigma_t (-(\delta \omega)^2} \int_{T_0}^{\infty} d T_{int}   \left[sgn(T_{int}-T_0 )-sgn(T_{int}-T_1) \right]\nonumber\\
 & &=(T_1-T_0)  2 \pi P_{\gamma_1}/V_{\gamma}  \int d \delta \omega  e^{-\sigma_t (\delta \omega)^2} = (T_1-T_0)  2 \pi P_{\gamma_1}/V_{\gamma}  \sqrt{\frac{\pi}{\sigma_t}}. \nonumber \\
\end{eqnarray}
The time and momentum are factorized.
 
(2-2) The  boundary  term,  $|B|^2$. 

We have the integral for a not-large $\sigma_t$,  using the variable $\xi=\sqrt{\frac{1}{\sigma_t}}(T_{int}-T_0) $,  and
 $\int_{T_0}^{\infty} dT_{int}  e^{  -\frac{1}{\sigma_t}(T_{int}-T_0)^2   } =\sqrt{\sigma_t}  \int _{0}^{\infty} d \xi e^{-\xi^2} =\sqrt{\sigma_t} \frac{\sqrt {\pi}}{2} $, 
as 
\begin{eqnarray}
\int dX_{\gamma_2,L} 2 \pi P_{\gamma_2}  d P_{\gamma_2}|B|^2&= &(\frac{\sqrt{2\sigma_t}}{2\sqrt \pi})^2\int_{T_0}^{T_1} dT_{int}   e^{  -\frac{1}{\sigma_t}(T_{int}-T_0)^2   } \int 2 \pi P_{\gamma_2}  d P_{\gamma_2}\frac{1}{\sigma_t^2(\delta \bar \omega )^2+(T_{int}-T_0)^2}  \nonumber \\
&= & P_{\gamma_1} \frac{1}{V_{\gamma}} \log{|T_{int}-T_0|}|_{\delta T}^{T_1 }+ \Delta, 
\end{eqnarray}
where $\Delta$ is a small correction, and  $\delta T$ is a short cut off time interval derived from length, of atomic size, or a wave length. For a large $\sigma_t$  of  $\sqrt \sigma_t \gg T_1-T_0$, $ \int_{T_0}^{T_1} dT_{int}  e^{  -\frac{1}{\sigma_t}(T_{int}-T_0)^2   } =(T_1-T_0)$,  
\begin{eqnarray}
\int dX_{\gamma_2,L}2 \pi P_{\gamma_2}  d P_{\gamma_2}|B|^2&= &(\frac{\sqrt{2\sigma_t}}{2\sqrt \pi})^2\int_{T_0}^{T_1} dT_{int}   e^{  -\frac{1}{\sigma_t}(T_{int}-T_0)^2   } \int 2 \pi P_{\gamma_2}  d P_{\gamma_2}\frac{1}{\sigma_t^2(\delta \bar \omega )^2+(T_{int}-T_0)^2}  \nonumber \\
&= &(\frac{\sqrt{2\sigma_t}}{2\sqrt \pi})^2\int_{T_0}^{T_1} dT_{int}    2 \pi P_{\gamma_1} \frac{1}{\sigma_t V_{\gamma}} \frac{1}{|T_{int}-T_0|}= P_{\gamma_1} \frac{1}{V_{\gamma}} \log{|T_{int}-T_0|}|_{\delta T}^{T_1 }.\nonumber \\
\end{eqnarray}
 $\Delta$ vanishes.

\section{ Understanding  blue sky is impossible with the standard formula }
The classical Rayleigh scattering was derived either  from Maxwell theory or from the Fermi's golden rule in the  quantum mechanics.  These agree with each others and have qualitatively correct  energy spctra  inversly proportional  to $E^{4}$. However, the absolute strength of diffusion component is much lower and  not in agreement with the observations even if higher order correction is included. Higher order correction  is  much weaker and neglibgible. \cite{higher_order}
We compare the standard calculations with the observations. \cite{Born and Wolf, Asano}

Optical depth is computed from the  Rayleigh cross section 
\begin{eqnarray}
& &\sigma=\frac{128\pi^5}{3\lambda^4}\alpha^2 \label{Rayleigh_c}\\
& &\alpha=\frac{m_r^2-1}{ 4 \pi N}
\end{eqnarray}
where $\alpha$ is the polarization and $N$ is the Avogadoro number, as 
\begin{eqnarray}
& &\tau_R(\lambda)=\int_0^{\infty} \sigma^R(\lambda) N(z) dz =\sigma^R(\lambda) \int_0^{\infty} dzN(z) \nonumber\\
& &\tau(\lambda,p_0)=0.008569 \lambda^{-4}(1+O(\lambda^{-2}) (\delta=0.031)\nonumber \\
& &\tau(\lambda,p)= \frac{p}{p_0} \tau(\lambda,p_0), 
\end{eqnarray}
where $p_0$ is the pressure of the atmosphere at $h=0$. 
$p_0=1013hPa$ and $ \lambda \text{ in unit of }(\mu \text{M})$
For $\lambda=0.5 \mu$M  
\begin{eqnarray}
& &\tau(\lambda,p_0)=0.008569 \times 2^4 (1+O(\lambda^{-2}) (\delta=0.031) =0.137 \label{depth_rayleigh}\nonumber\\
& &\tau(\lambda,p)= \frac{p}{p_0} 0.137  
\end{eqnarray}

The  optical depth Eq.$(\ref{depth_rayleigh})$ is around $0.14$  at $h=0$ and $0.041$  at $h=10$ kM, where $ P=P_0 \times 0.3$. These values show that the fraction of $85$ per cent at the earth surface and  $96$ per cent at $h=10$ kM, which is the height of jet planes, of the solar light is not disturbed.  If the strength of diffused solar light at $h=10$ kM  would have been  $4$ per cent of the incoming flux, the sky at off axis direction  must be dark. However,  we know from dayly  experiences from jet planes that the strength of blue color of the sky at $h=10$ kM is the same as the ground $h=0$.  Observation does not agree  with the values obtained from the clasical formula or  Fermi's goldedn rule on Rayleigh scattering.  Jet plane started in 1952, much later than Rayleigh's work.

\end{document}